\documentstyle[preprint,aps,prc,floats,psfig]{revtex}
\draft
\renewcommand{\thefootnote}{\arabic{footnote}}
\def\gev{\rm GeV}
\def\fbi{\rm fb^{-1}}

\begin{document}

\draft
\renewcommand{\thefootnote}{\arabic{footnote}}

\preprint{
\vbox{\hbox{\bf MADPH-99-1147}
      \hbox{\bf hep-ph/0006223}
      \hbox{June, 2000}}}

\title{\bf $\tan\beta$ Determination From Heavy Higgs Boson Production\\ 
at Linear Colliders}
\author{V. Barger,\footnote{barger@oriole.physics.wisc.edu}
T. Han\footnote{than@egret.physics.wisc.edu}
and J. Jiang\footnote{jiang@pheno.physics.wisc.edu}}
\address{Department of Physics, University of Wisconsin,\\ 
1150 University Avenue, Madison, WI 53706, USA}

\maketitle


\begin{abstract}
We study the production at future $e^+e^-$ linear colliders of the heavy 
neutral Higgs bosons $H$ and $A$ of the minimal supersymmetric 
standard model in association with top and bottom quarks. 
The cross sections have a strong dependence on the parameter 
$\tan\beta$, and thus provide a good way to determine it.  
At a linear collider with $\sqrt s = 0.5-1$ TeV and expected 
integrated luminosities, we find significant sensitivities for 
determining $\tan\beta$. In the Supergravity scenario, the
sensitivity is particularly strong for $\tan\beta \agt 10$,
reaching a $15\%$ or better measurement. 
In the general MSSM scenario, the interplay
between the $4b$ and $4t$ channels results in a good
determination for $\tan\beta \alt 10$, while the sensitivity 
is weakened for higher values of $\tan\beta$.
\end{abstract}

\newpage

\section{Introduction}

One of the most promising avenues for physics beyond the standard model (SM)  
is supersymmetry (SUSY) \cite{susy}, since it can provide a fundamental 
understanding of electroweak symmetry breaking (EWSB)
and it allows unification of the electroweak  
and strong interactions at a grand unified scale \cite{susyreview}. 
Because of its great theoretical  attraction, 
extensive phenomenological work continues to explore the ways for  
discovery and precision study of supersymmetric particles at present and  
future colliders. 

Most of these investigations are directed to the Minimal 
Supersymmetric standard model (MSSM), 
which has the minimal new particle content \cite{susy}. 
The MSSM contains two Higgs doublets which develop vacuum expectation  
values $\left<H_1\right> = v_1/\sqrt 2$ and
       $\left<H_2\right> = v_2/\sqrt 2$ 
that break the $\rm SU(2)\times U(1)$ 
gauge symmetry spontaneously \cite{higgs}. There are 5  
physical Higgs bosons in the MSSM: two CP-even states $h$ and $H$, a CP-odd  
state $A$, and two charged states $H^\pm$; the lightest Higgs boson is $h$.

The ratio $v_2/v_1=\tan\beta$ is a critical parameter of the MSSM:
It characterizes the relative fraction that the two Higgs doublets 
contribute to the EWSB, and it enters all sectors of the theory. 
The interactions of both the SUSY particles and the Higgs bosons depend on  
$\tan\beta$, and the relations of SUSY particle masses to the soft symmetry  
breaking parameters of supersymmetry involve $\tan\beta$\cite{higgs}.
A measurement of $\tan\beta$ from one sector will thereby 
allow predictions or tests in other sectors. 
The renormalization group evolution of the Yukawa 
couplings from the unification scale to the electroweak scale 
are sensitive to the value of $\tan\beta$. The large top  
quark mass can naturally be explained with $m_b-m_\tau$ unification 
as a quasi-infrared fixed point of the top Yukawa coupling if 
$\tan\beta\simeq 1.8$ or $\tan\beta \simeq 56$ \cite{BBO}.  
The possibility of SO(10) Yukawa unification  
requires the high $\tan\beta$ solution \cite{so(10)}. 
The predicted mass of the lightest SUSY Higgs boson also 
depends on $\tan\beta$, with $m_h \sim 105~$GeV at  
$\tan\beta\simeq 1.8$ and $m_h \sim 120$~GeV at 
$\tan\beta\agt20$ \cite{deltamh}.

Because of the significance of $\tan\beta$ for the theory and phenomenology  
of the MSSM, it is important to find processes in which $\tan\beta$ can be  
best determined. Some regions of the MSSM parameter space have 
been excluded at LEP2 \cite{lep2} due to the lower bound on the lightest 
Higgs boson mass ($m_h$), particularly at low $\tan\beta$ near 1. 
Much of the parameter space remains to be explored at the upgraded
Tevatron \cite{tev,run2}, the LHC \cite{LHC,lhctanb,baer}, the
future linear colliders \cite{nlctanb,Nojiri,hh,feng,han,lctanb} and 
muon colliders \cite{muon}.
The $\tan\beta$ constraints that may be obtained from
$m_h$ via radiative corrections \cite{deltamh},
or from precision electroweak measurements \cite{yamada}, or from 
SUSY particle production usually depend also on other SUSY parameters.
Furthermore, measurements of $\sin\beta$ or $\cos\beta$ via other
SUSY processes without directly involving Higgs bosons do 
not accurately determine large $\tan\beta$ values \cite{han}. For general
SUSY Higgs phenomenology, we refer the readers to  
reviews \cite{snow}.

The Higgs couplings of $H, A, H^\pm$ to heavy quarks are given by
\begin{eqnarray}
&&A\bar tt :  
\frac{-gm_t}{2m_W}\cot\beta\ {\gamma}_5 \qquad \qquad\qquad\qquad\ 
A\bar bb :  \frac{-gm_b}{2m_W}\tan\beta\ {\gamma}_5 \label{equ:Aff}\\ 
&&H\bar tt :  \frac{-igm_t}{2m_W}\frac{\sin\alpha}{\sin\beta} \approx
       \frac{igm_t}{2m_W}{\cot\beta}\quad \qquad
H\bar bb :  \frac{-igm_b}{2m_W}\frac{\cos\alpha}{\cos\beta} \approx
       \frac{-igm_b}{2m_W}{\tan\beta}\label{equ:Hff}\\ 
&&H^+\bar{t}b :  \frac{igV_{td}}{2\sqrt{2}m_W}[(m_b\tan\beta+m_t\cot\beta)
+(m_b\tan\beta-m_t\cot\beta){\gamma}_5],
\label{equ:Htb}
\end{eqnarray}
where the decoupling limit $M_A\gg M_Z$ has been assumed
for the approximate forms of $H\bar tt,H\bar bb$. In this
limit, the lightest Higgs boson $h$ becomes SM-like and its
couplings are insensitive to SUSY parameters. For the
$H, A, H^\pm$ Higgs bosons,  
$\tan\beta$ is essentially the unique parameter for Higgs-heavy 
quark couplings. This suggests that studies of the
associated production of the Higgs bosons and heavy quarks
may effectively probe the $\tan\beta$ parameter.

Heavy Higgs boson production at future $e^+e^-$ colliders
was discussed in Ref.~\cite{hh}.
In a recent study Feng and Moroi\cite{feng} evaluated the prospects 
for determining $\tan\beta$ in $e^+e^-$ collisions with 
$\sqrt s = 0.5$ and 1~TeV c.m.~energy via the processes 
\begin{equation}
e^+e^-\to Zh,\ AH,\ t\bar bH^-\quad {\rm and}\quad \bar t b H^+.
\end{equation} 
The primary channel in their study involves the $H^+\bar t b$ coupling.
They found that the strong dependence of heavy Higgs branching fractions 
on $\tan\beta$ allows stringent constraints to be placed for moderate 
$\tan\beta$\cite{feng} in the MSSM.  
In the present paper, we report results of a complementary study of 
the associated production of a neutral Higgs boson 
and heavy quarks
\begin{equation}
e^+e^-\to H\bar tt,\ H\bar bb,\ A\bar tt,\quad {\rm and}\quad A\bar bb. 
\label{hqq}
\end{equation}
These processes involve $\bar t t$ and $\bar b b$ production
separately and are thereby expected to be complementary for 
low and high values of $\tan\beta$. We study the sensitivity to
probe $\tan\beta$ in two scenarios: the minimal
Supergravity model (mSUGRA) and the MSSM.

The paper is organized as follows: We present the Higgs decay
branching fractions and the cross sections for the associated 
production of the Higgs bosons and heavy quarks in Sec.~II. 
We analyze the sensitivity to determine
the value of $\tan\beta$ at future linear colliders in Sec.~III.
We discuss our results, make some general remarks and conclude
in Sec.~IV.

\section{Neutral Higgs Production}

\subsection{Input Parameters}

The mass matrix of the CP-even Higgs bosons of the MSSM is given by
\begin{equation}
{\cal M}^2 = \left(
\begin{array}{cc}
m_A^2 \sin^2\beta + m_Z^2 \cos^2\beta & -(m_A^2+m_Z^2) \sin\beta \cos\beta\\
-(m_A^2+m_Z^2) \sin\beta \cos\beta& m_A^2 \cos^2 \beta + m_Z^2 \sin^2\beta
\end{array} \right) + \Delta {\cal M}^2 \,,
\end{equation}
where $\Delta {\cal M}^2$ represents the radiative corrections. 
At tree level, the input parameters are $m_A$ and $\tan\beta$,
and the phenomenology is relatively easy to analyze.
The radiative corrections may be substantial in this CP-even sector 
and they are dependent on other SUSY parameters, 
especially on the masses and couplings/mixings of the top quark
and heavy scalar quarks (squarks) \cite{deltamh}. In a general MSSM analysis 
the parameters required as input are 
\begin{equation}
m_Q,\quad m_U,\quad m_D,\quad 
M_1,\quad M_2,\quad A_U,\quad A_D\quad {\rm and}\quad \mu, 
\end{equation}
where $m_Q$ is the soft SUSY breaking mass parameter of left-handed 
stop (where only the heavy third generation parameters are relevant), 
$m_U\ (m_D)$ the SUSY breaking mass parameter of right-handed 
stop (sbottom), $M_1, M_2$ the gaugino masses,
$A_U\ (A_D)$ the stop trilinear soft breaking term, and $\mu$ the 
Higgs mixing parameter.  The large parameter space involved 
makes phenomenological studies difficult.  On the other hand,
once a precision measurement is made in the Higgs sector in
future collider experiments, we would expect to learn more about 
the SUSY sector due to the radiative relations among the
physical SUSY masses.
Instead of exploring the large space of the MSSM soft parameters,
we focus on the following two scenarios for illustration.

\noindent
{\bf \underline{mSUGRA}}
  
Motivated by the mSUGRA model and 
the requirements of radiatively generated electroweak
symmetry breaking (EWSB), we relate 
the soft SUSY breaking parameters to the common scalar,
fermion and trilinear parameters 
\begin{equation}
m_0,\quad m_{1/2}\quad {\rm and}\quad A_0
\end{equation} 
at the grand unified scale. 
For specific choices of $\tan\beta$ the results depend on the sign of 
$\mu$. The $\mu > 0$ sign is less constrained by 
$b \to s\gamma$ decay \cite{btosg}
and we adopt this convention in our analysis.  

We make use of the ISAJET package \cite{isajet} to determine
the SUSY masses and couplings from the GUT scale input parameters. 
The Higgs mass eigenvalues are among the outputs of this program.  
These values agree 
with the corresponding results from the code of Ref.~\cite{hmass} 
to a precision $\alt 0.3\%$. The soft-supersymmetry-breaking parameters 
are evolved according to 
renormalization group (RG) equations \cite{BBO,RGE,SUGRA}.  
For our illustrations we make the parameter choice
\begin{equation}
m_0 = 250\ {\gev}, \quad m_{1/2} = 150\ {\gev}, \quad A_0 = -300\ {\gev},
\label{msugra}
\end{equation}
along with the positive sign of $\mu$.  
The magnitude of $\mu$ is fixed in terms of $M^{}_Z$ through 
the radiately generated EWSB. For three representative $\tan\beta$ 
values, the mass eigenvalues of Higgs bosons and SUSY soft-breaking 
terms are listed in Table~\ref{para}.
For charginos and neutralinos, we list only the masses of the 
lightest ones. In fact, our choice of the above
parameters is somewhat conservative in exploring the SUSY Higgs sector.
A large $m_0$ results in heavy $H,A,H^\pm$. Consequently it
leads to the ``decoupling limit'' \cite{hhaber} so that
the lightest Higgs boson $h$ becomes SM-like and thus insensitive
to the SUSY parameters. 
\begin{table}[tb]
\begin{center}
\begin{tabular}{ccccccccccc}
$\tan\beta$ & $m_A$ & $m_H$ & $m_{\chi^\pm}$ & $m_{\chi^0}$ & $\mu$ &
$m_Q$ & $m_U$ &
$m_D$ & $A_U$ & $A_D$ \\
\hline
3  & 434 & 438 & 105 &56 & 315 & 360 & 265 & 411 & $-330$ & $-693$\\
\hline
10 & 373 & 373 & 110 &57 & 274 & 359 & 272 & 406 & $-370$ & $-689$\\
\hline
30 & 273 & 273 & 112 &59 & 264 & 337 & 276 & 364 & $-354$ & $-581$\\
\end{tabular}
\vskip 0.3cm
\caption{ISAJET output parameters based on the mSUGRA input
in Eq.~(\ref{msugra}).
\label{para}}
\vskip -1cm
\end{center}
\end{table}

\noindent
{\bf \underline{MSSM}}

We also perform the same study in the MSSM scenario, in which 
$\tan\beta$ as well as the masses of the Higgs bosons (determined
by $m_A$) are all free parameters to explore.
The choice of other input parameters is as follows
\begin{eqnarray}
&&\mu = 272\ {\gev}, \quad m_Q = 356\ {\gev}, \quad m_U = 273\ {\gev}, \quad\\
&&m_D = 400\ {\gev},  \quad A_U = -369\ {\gev}, \quad A_D = -672\, {\gev} 
\quad\\
&&m_{\chi^\pm} = 111\ {\gev}, \quad m_{\chi^0} = 59\ {\gev}.
\label{mssmp}
\end{eqnarray}
These soft SUSY breaking parameters are similar to mSUGRA parameters 
with $\tan\beta\approx 15$. In particular we study two cases 
with $m_A^{} = 200$ GeV and 400 GeV, while $m_H^{}$
is nearly degenerate with $m_A^{}$. These choices represent
the kinematical situations for $A, H$ to be below and above $t\bar t$
threshold.

\begin{figure}[tbh]
\centerline{\psfig{file=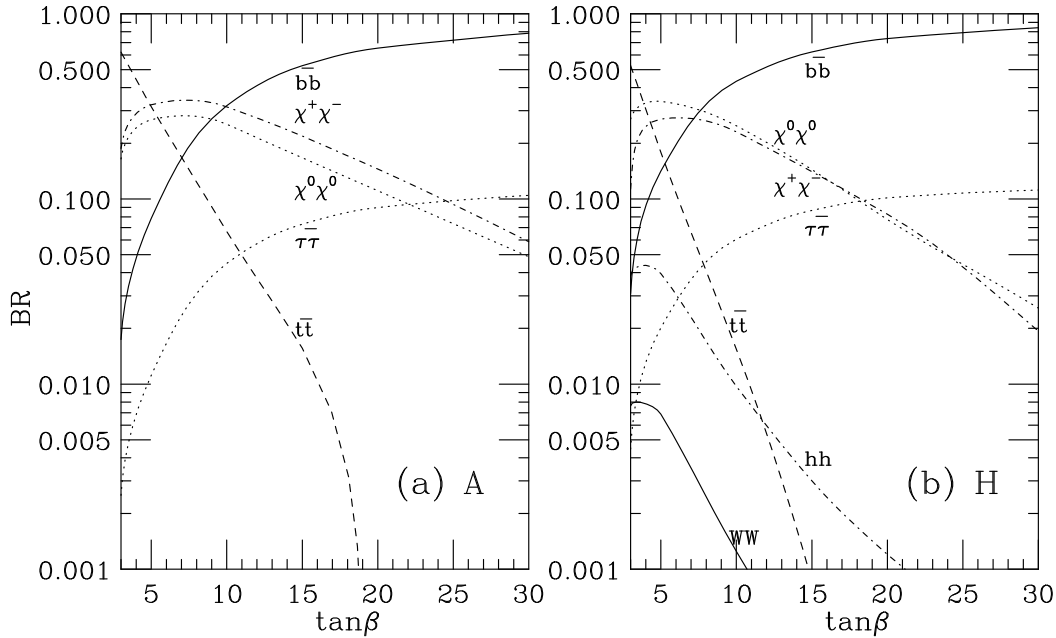,width=5.5in,angle=90}}
\bigskip
\caption{mSUGRA: Leading branching fractions of 
decays (a) of $A$ and (b) of $H$ versus $\tan\beta$.
\label{fig:Abr}}
\end{figure}
	  
\subsection{Branching Fractions}

We use the program provided in Ref.~\cite{hmass} for obtaining 
the branching fractions for the Higgs boson decay. In this 
program all kinematically allowed decay channels in MSSM are 
included and RG improved values of Higgs masses and couplings 
with the main NLO corrections \cite{deltamh} are implemented.

In Fig.~\ref{fig:Abr} we plot the branching fraction of the decays
(a) for $A$ and (b) for $H$ versus $\tan\beta$ in mSUGRA.  
As $\tan\beta$ increases, the branching fractions 
of $A$ and $H$ decay into $t\bar{t}$ drop rapidly and the decays into 
$b\bar{b}$ increase dramatically. The branching fractions into chargino 
and neutralino pairs peak at intermediate values $\tan\beta\sim 5$ 
and can be as large as 30\%.  
Branching fractions of $H$ decay into $hh$ and $WW$ 
are also shown in Fig.~\ref{fig:Abr} (b) for comparison.
With this strong dependence of the branching fractions on $\tan\beta$, 
we expect neutral Higgs production channels to be useful in 
determining the value of $\tan\beta$. In particular, it is 
interesting to note the complementarity between $t\bar t$
and $b\bar b$ modes for small and large values of $\tan\beta$.

\begin{figure}[tbh]
\centerline{\psfig{file=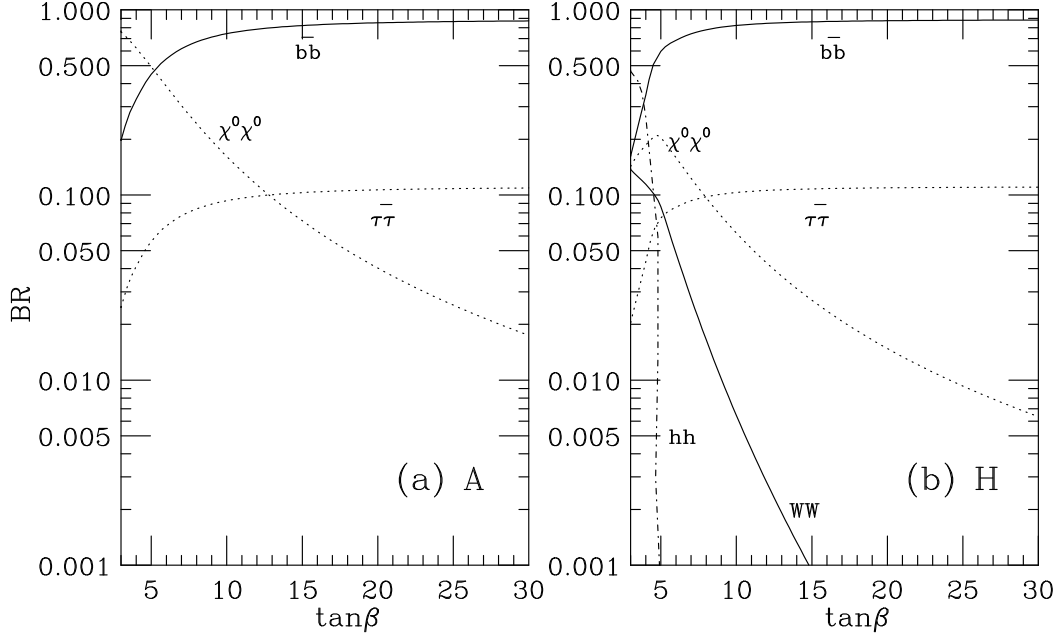,width=5.5in,angle=90}}
\caption{MSSM: 
Leading branching fractions of decays with $m_A = 200$ GeV
 (a) of $A$ and (b) of $H$ versus $\tan\beta$.
\label{fig:br200}}
\end{figure}

\begin{figure}[tbh]
\centerline{\psfig{file=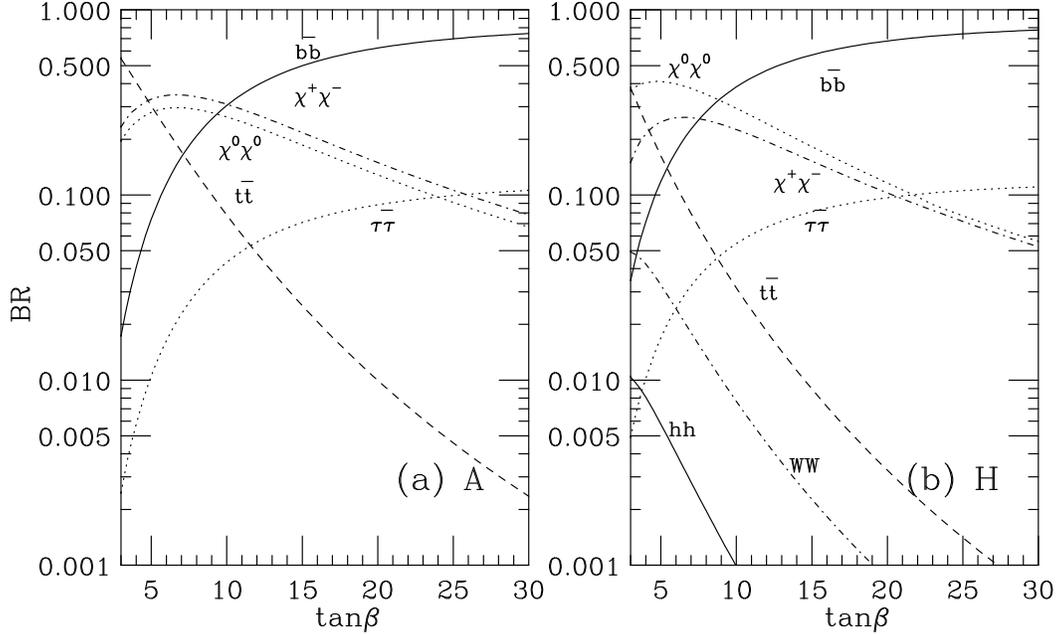,width=5.5in,angle=90}}
\caption{MSSM: 
Leading branching fractions of decays with $m_A=400$ GeV 
 (a) of $A$ and (b) of $H$
versus $\tan\beta$.
\label{fig:br400}}
\end{figure}

In the MSSM scenario, for the case of $m_A$ = 200 GeV,
Figure \ref{fig:br200} shows the branching fraction 
of the decays of $A$ and $H$ versus $\tan\beta$. Note that
the $t\bar t$ channel is not open. For the case of
$m_A$ = 400 GeV, the corresponding branching fractions are shown in
Fig.~\ref{fig:br400}. We see that Figs.~\ref{fig:Abr} and 
\ref{fig:br400} are alike due to similar kinematical
thresholds.

\subsection{Cross Sections and Final State Signature}

\begin{figure}[tbh]
\centerline{\psfig{file=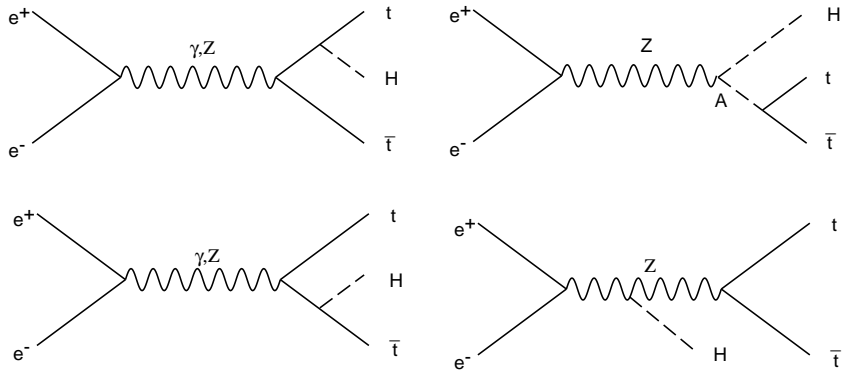,width=4.5in}}
\caption{Feynman diagrams contributing to $e^+e^- \to Ht\bar{t}$.  The
diagrams for $e^+e^- \to At\bar{t}$ are similar, except that the last 
diagram above is absent.  
\label{fig:Feym}}
\end{figure}

As a representative example of the processes in Eq.~(\ref{hqq}), 
the tree-level Feynman diagrams for 
$e^+e^- \to Ht\bar{t}$ are shown in Fig.~\ref{fig:Feym}.  
For the other processes, we simply need to replace $H$ with $A$, or/and 
$t\bar{t}$ with $b\bar{b}$. The last diagram in Fig.~\ref{fig:Feym} 
involving the $ZZH$ coupling
is unique to the process which has $H$ in the final state.  
We have included both the diagrams of Higgs radiation off
a heavy quark ($Ht\bar{t}$) and Higgs decay ($HA\to Ht\bar t$).
It is important to note that the $H$ ($A$) decay processes 
are sensitive to $\tan\beta$ only
when the branching fractions vary rapidly. The $H,A\to b\bar b$
branching fractions gradually approach unity
at large $\tan\beta$, and the dependence on $\tan\beta$ is 
thus reduced here. On the other hand, diagrams with $H$ ($A$) 
radiation off a heavy quark 
typically have a quadratic dependence on $\tan\beta$, and are thus
quite sensitive to $\tan\beta$.

\begin{figure}[tbh]
\centerline{\psfig{file=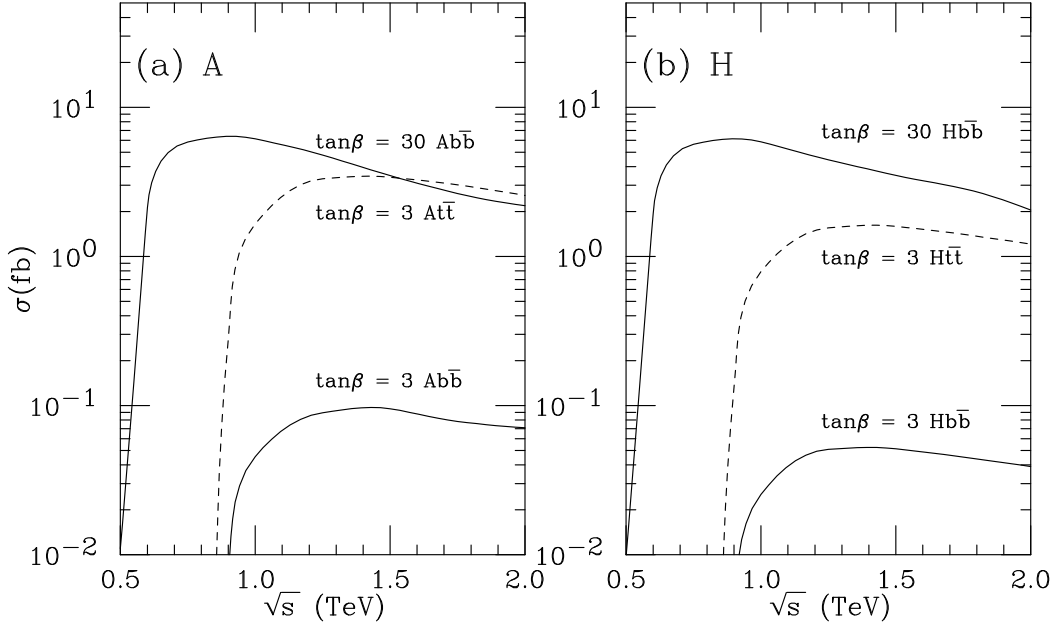,width=5.5in,angle=90}}
\caption{mSUGRA: Total Higgs production cross sections versus 
the center of mass
energy with $\tan\beta$ = 3 and 30
(a) for $e^+e^- \to At\bar{t}$ (dashes)
and $Ab\bar{b}$ (solid) and  (b) for $e^+e^- \to Ht\bar{t}$ (dashes) 
and $Hb\bar{b}$ (solid).  
\label{fig:tot}}
\end{figure}

\begin{figure}[tbh]
\centerline{\psfig{file=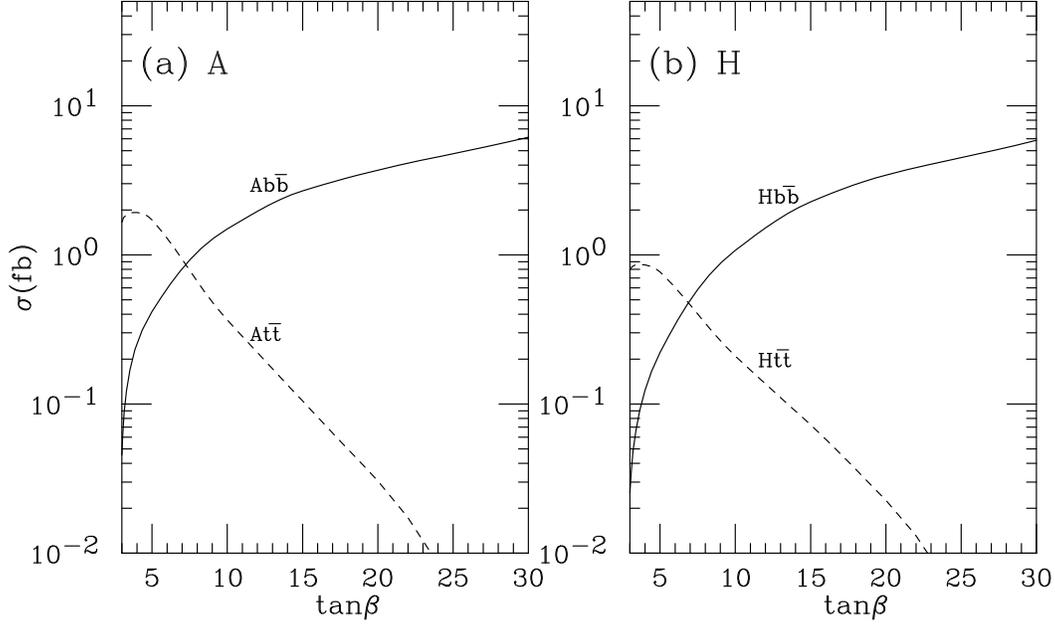,width=5.5in,angle=90}}
\caption{mSUGRA: Total Higgs production cross sections 
versus $\tan\beta$ 
at $\sqrt{s} = 1$ TeV (a) for $e^+e^- \to At\bar{t}$ (dashes)
and $Ab\bar{b}$ (solid) and  (b) for $e^+e^- \to Ht\bar{t}$ (dashes) 
and $Hb\bar{b}$ (solid).  
\label{fig:totTB}}
\end{figure}

\begin{figure}[tbh]
\centerline{\psfig{file=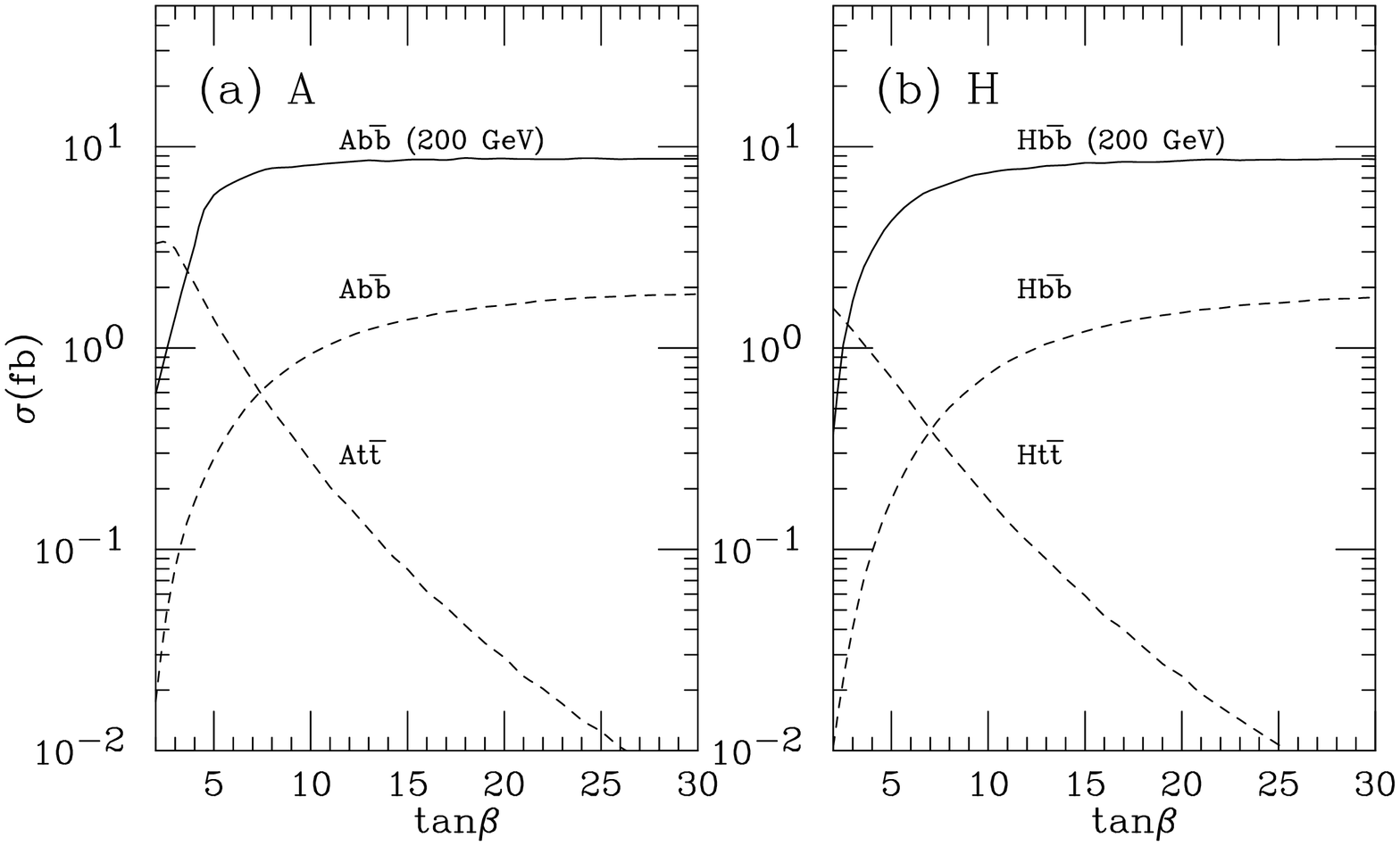,width=5.5in,angle=90}}
\caption{MSSM: Total Higgs production cross sections versus 
$\tan\beta$ (a) for $e^+e^- \to At\bar{t}$ and $Ab\bar{b}$,
and (b) for $e^+e^- \to Ht\bar{t}$ and $Hb\bar{b}$.
The solid curves are for $m_A$ = 200 GeV at
$\sqrt{s} = 500$ GeV; the dashes are for $m_A$ = 400 GeV 
at $\sqrt{s} = 1$ TeV.
\label{fig:h2qmssm}}
\end{figure}

Figure~\ref{fig:tot} shows the calculated total cross sections of the
processes $e^+e^- \to A(H)t\bar{t}, A(H)b\bar{b}$ versus the center 
of mass energy ($\sqrt{s}$) for $\tan\beta$ = 3 and 30 in the
mSUGRA scenario. The cross sections for $Ab\bar b, At\bar t$ 
can be typically of $0.1 - 10$ fb for this range of $\tan\beta$ 
at linear collider energies of $0.5-2$ TeV. The maximum
rate is reached at a c.m.~energy about 300 GeV or so
above the $At\bar t$ threshold. 
Note the different mass thresholds in this figure
for the two values of $\tan\beta$, as given by the masses in 
Table~\ref{para}.
For the heavy Higgs bosons under consideration, we
concentrate on a collider energy $\sqrt s\sim 1$ TeV.
We plot the cross sections versus $\tan\beta$ in 
Fig.~\ref{fig:totTB} again in the mSUGRA scenario. 
At low $\tan\beta$ the associated production of $A$ with $t\bar{t}$ 
is dominant but this channel is greatly suppressed at large $\tan\beta$ 
values.  On the other hand the production of $A$ in association with 
$b\bar{b}$ is small at low $\tan\beta$ and increases rapidly with 
$\tan\beta$. Figures \ref{fig:tot} and \ref{fig:totTB} show that
associated production of $H$ with $t\bar{t}$ or $b\bar{b}$ has similar 
characteristics to $A$ production.
Figure \ref{fig:h2qmssm} shows the cross sections similar 
to Fig.~\ref{fig:totTB} but in the MSSM scenario for  
cases: $m_A$ = 200 GeV at $\sqrt{s} = 500$ GeV (solid)
and 400 GeV at $\sqrt{s} = 1$ TeV (dashes).

\begin{figure}[tbh]
\centerline{\psfig{file=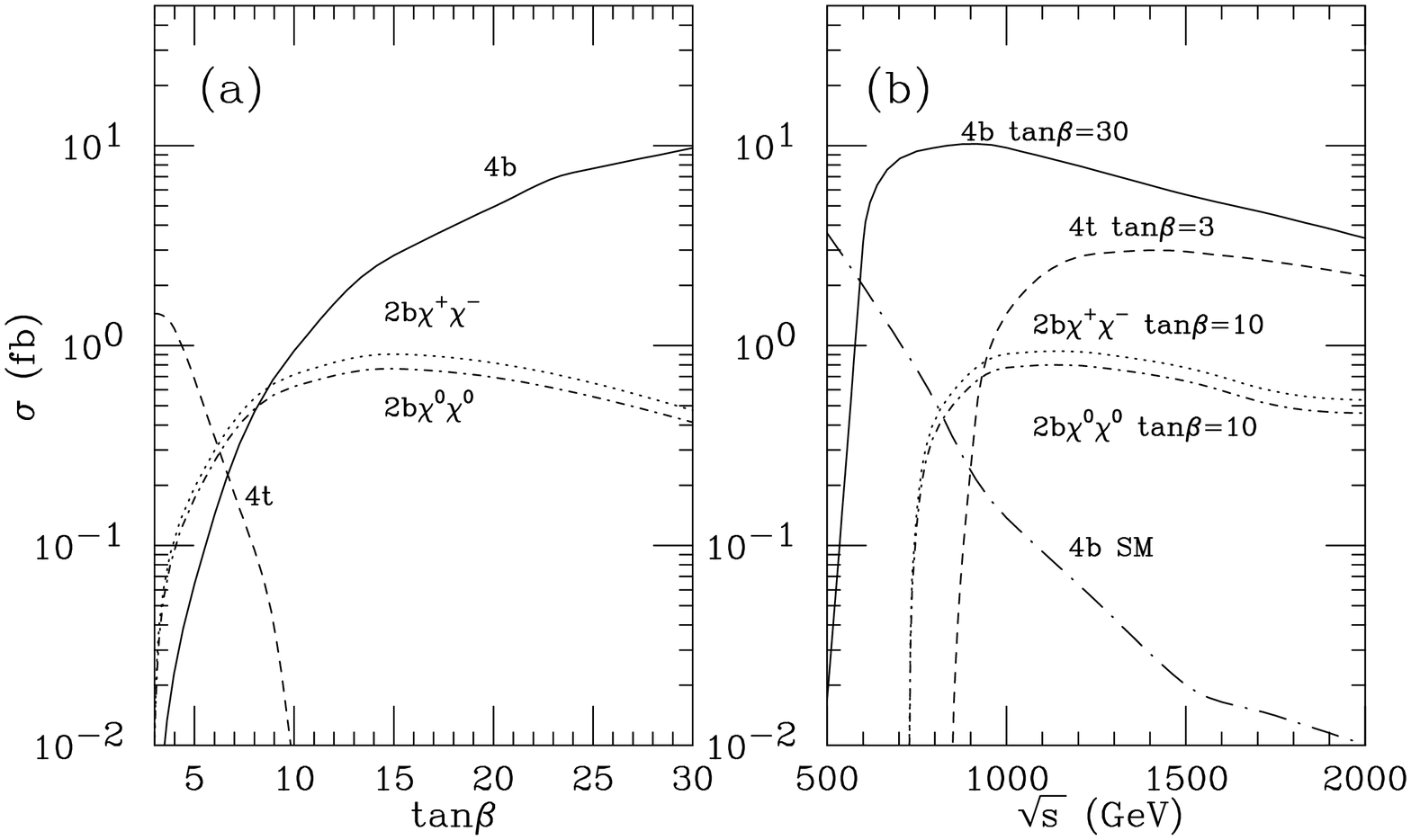,width=5.5in,angle=90}}
\caption{mSUGRA: Total cross sections with different final states
including $A,H$ decays (a) versus $\tan\beta$ at $\sqrt s=1$ TeV,
and (b) versus $\sqrt s$ for representative values of
$\tan\beta =3,10$ and 30. The SM expectation of $4b$ production
is also included for comparison.
\label{fig:signals}}
\end{figure}

Concerning the final state signature with the
$A (H)$ decays, we notice that at low $\tan\beta$, 
both the production cross section for $At\bar{t}$ $(Ht\bar{t})$ 
and the branching fraction for $A$ ($H$) decay into 
$t\bar{t}$ are large as a result of the typical ($\cot\beta)^4$ 
enhancement. The $e^+e^- \to t\bar{t}t\bar{t}$ signal
is dominant at low $\tan\beta$ but at high $\tan\beta$, 
$e^+e^- \to b\bar{b}b\bar{b}$ becomes dominant because 
of the $(\tan\beta)^4$ enhancement.
For intermediate values of $\tan\beta \sim 5$, the SUSY
decay modes, such as $A,H\to \chi^+\chi^-$ and 
$\chi^0\chi^0$ can
be more important. We show in Fig.~\ref{fig:signals}(a) the
total cross sections at $\sqrt s=1$ TeV versus $\tan\beta$
including the different final states.
The complementarity of the three final states in different
range of $\tan\beta$ can be seen in this figure.
Figure~\ref{fig:signals}(b) again shows the contribution of these
final states for $\tan\beta$ values where they are most important: 
$4b$ for $\tan\beta =30$, $b\bar{b}\chi^{\pm}\chi^{\mp}$ or 
$b\bar{b}\chi^0\chi^0$ for $\tan\beta =10$, and $4t$ for
$\tan\beta =3$. The $4b$ standard model background
is also shown by the dot-dashed curve.

\begin{figure}[tbh]
\centerline{\psfig{file=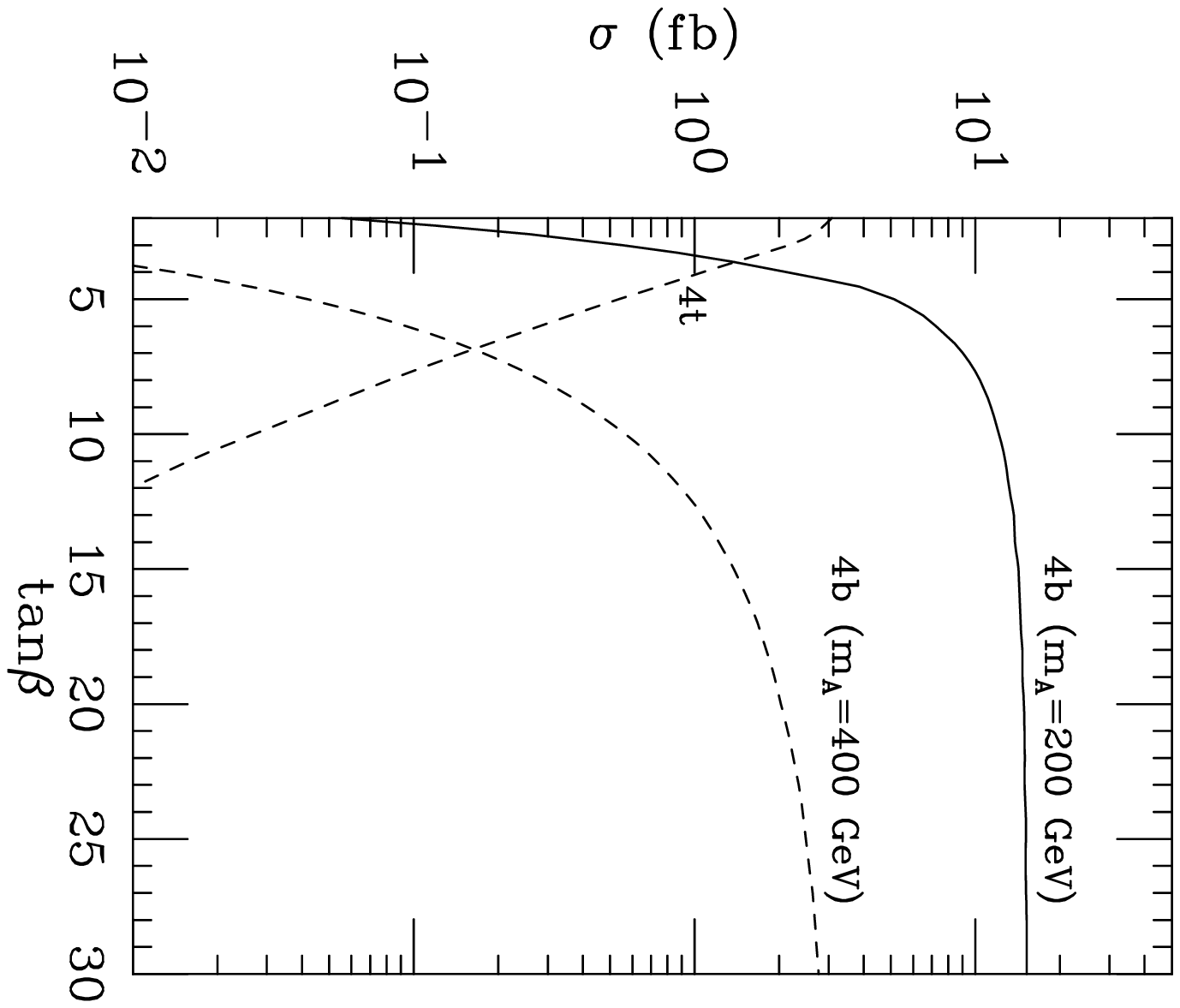,width=3in,angle=90}}
\caption{MSSM: Total cross sections with $4t$ and $4b$
final states including $A,H$ decays versus $\tan\beta$.
The solid curve is for $m_A$ = 200 GeV at $\sqrt{s} = 500$ GeV; 
the dashes are for $m_A$ = 400 GeV at $\sqrt{s} = 1$ TeV.
\label{fig:e4qmssm}}
\end{figure}

In the MSSM scenario, similar curves for the $4t$ and $4b$
final state signals are shown in Fig.~\ref{fig:e4qmssm} 
for two cases: $m_A = 200$ GeV at $\sqrt{s} = 0.5$ TeV 
and $m_A = 400$ GeV at $\sqrt{s} = 1$ TeV.

\subsection{Background} 

The most robust channels, $b\bar b b\bar b$ and $t\bar t t\bar t$, 
from neutral Higgs production have rather small SM backgrounds. 
The SM expectation for $e^+e^-\to b\bar b b\bar b$ 
production is shown in Fig.~\ref{fig:signals}(b). The cross section
decreases with increasing $\sqrt{s}$ as $(1/\sqrt{s})^2$.
At $\sqrt{s} = 1$ TeV, the $4b$ background is only 0.1 fb, 
much smaller than the signal rate at large $\tan\beta$.
The SM cross section for $e^+e^-\to t\bar{t}t\bar{t}$ 
is smaller than $10^{-3}$ fb at $\sqrt{s} = 1$ TeV and
 thus is negligible.  The SM $4b$ background at $500$ GeV is
about $3.7$ fb.  We take this background into consideration
when we calculate the limits at $\sqrt{s} = 500$ GeV.
Since the SM backgrounds are small relative to the signals of interest, 
we do not need to impose sophisticated kinematical cuts 
and the signal rates are thereby better preserved.
The final states involving the
gauginos may have rather large SM backgrounds from
$b\bar b$, $t\bar t$, gauge boson production.  We neglect those 
channels in our evaluation.

\section{Analyses and results} 

As the parameter $\tan\beta$ is varied from small to intermediate
to large values, the dominant Higgs signal comes from the three 
channels $t\bar{t}t\bar{t}$, $b\bar{b}\chi\chi$, $b\bar{b}b\bar{b}$, 
respectively. Since the sizes of the signal cross sections depend 
sensitively on $\tan\beta$, a determination of $\tan\beta$ should be 
possible throughout $\tan\beta$ ranges where there are substantial signal 
event rates. In our analyses, we employ the $t\bar t t\bar t$ signal
at low $\tan\beta$ and the $b\bar b b\bar b$ signal
at large $\tan\beta$. For the intermediate $\tan\beta$ values, 
we combine these two channels. We do not include the channels with
gaugino final state in our consideration
 since the signatures would depend upon other
SUSY parameters such as the slepton and squark masses. We thus
regard the results of our analyses to be conservative.

We consider a $\sqrt{s} = 1$ TeV collider with three integrated 
luminosities of 50, 100 and 500 fb$^{-1}$.  After applying the 
geometrical cut 
\begin{equation}
\cos(\theta_b) < 0.9
\end{equation}
to the $4b$ signals, the 
total cross section is reduced to $23\%$, which we take as the 
geometrical efficiency.  Because of the low background cross 
section, we only need low purity of $b$-tagging; we assume a 
$b$-tagging efficiency, $\epsilon_b \approx 65\%$ \cite{snowmass}.  
Since $b$-quark flavors are conserved in the production process, 
we can relax the requirement to tag only three $b$-quarks, 
as is a standard practice. Then the efficiency of detecting 
$3b$ in a $4b$ sample is 
$4{\epsilon_b}^3-3{\epsilon_b}^4 \approx 56\%$.
For the $4t$ channel, although the event kinematics would be
more involved, the distinctive event topology compared to the SM
multi-jet backgrounds should allow a clear signal separation.
Nonetheless, we still require the identification of at least 
three $b$ quarks.  At a given $\tan\beta$ value, 
we multiply the total cross section of $4t$ or $4b$ channel 
with the geometrical efficiency, the $b$-tagging efficiency, and 
the integrated luminosity to get the signal event rate $N_S$.  
The statistical standard deviation is $\sigma = \sqrt{N_S}$.
In the presence of SM backgrounds, we similarly determine
the background event rate $N_B$.  We then take the conservative
estimate for the signal fluctuation
\begin{equation}
\sigma = \sqrt{N_S+N_B}.
\end{equation}  
For a $95\%$ confidence level (C.L.) cross section measurement, 
the range for the number of events is taken to be 
$N_S \pm 1.96\sigma$. The corresponding bounds on the signal
cross sections can be translated into allowed ranges 
$\Delta\tan\beta$ given by
\begin{equation}
\Delta\tan\beta = \tan\beta_\pm - \tan\beta,
\end{equation}
where $\tan\beta$ is determined from $N_S$ and $\tan\beta_\pm$
is determined from $N_S\pm1.96\sqrt{N_S+N_B}$.

\begin{figure}[tbh]
\centerline{\psfig{file=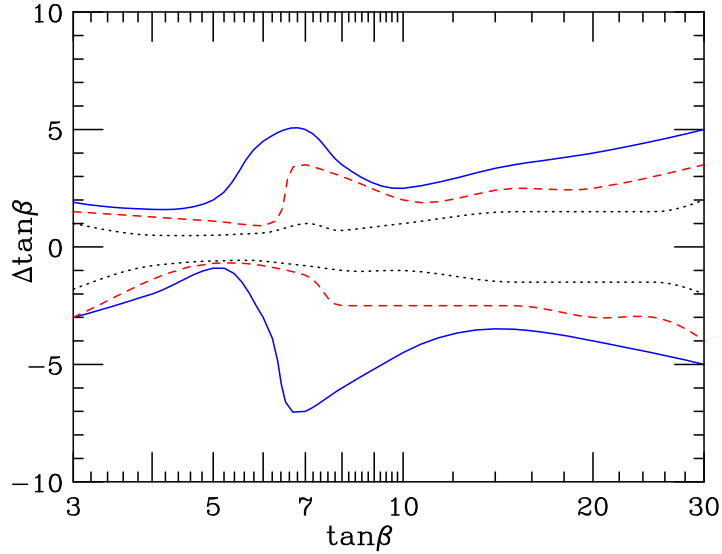,width=3.75in,angle=90}}
\caption{mSUGRA: Determination of $\tan\beta$ at $\sqrt{s}=$1 TeV 
combining both $A$ and $H$ channels;
$95\%$ C.L.~constraints on the $\tan\beta$ 
values are shown for 50 fb$^{-1}$ (solid), 
100 fb$^{-1}$ (dashes) and 500 fb$^{-1}$ (dotted).
\label{fig:const}}
\end{figure}

We first consider the mSUGRA scenario at a $\sqrt{s}=$1 TeV
linear collider. We combine both $A$ and $H$ channels.
In Fig.~\ref{fig:const}, the $95\%$ C.L.~constraints on
$\Delta\tan\beta$ for 50 fb$^{-1}$ (solid),
100 fb$^{-1}$ (dashes) and 500 fb$^{-1}$ (dotted) are shown 
versus $\tan\beta$. We find encouraging results for
the $\tan\beta$ determination. For instance, with a luminosity
of 100 fb$^{-1}$, $|\Delta\tan\beta|\approx 3$ or better can be 
reached at the low value of $\tan\beta$, mainly via the
$4t$ channel. At the high value 
of $\tan\beta$, 
$|\Delta\tan\beta|\approx 5$ can be reached,
 mainly via the $4b$ channel,
which is better than $15\%$ accuracy.
The slightly more difficult region is $\tan\beta\approx 6-7$,
where the $4t$ and $4b$ channels both yield smaller contributions.
We expect that the inclusion of the chargino channels would
improve the determination. Nevertheless, a good determination
has been seen for the whole $\tan\beta$ range of interest.

\begin{figure}[tbh]
\centerline{\psfig{file=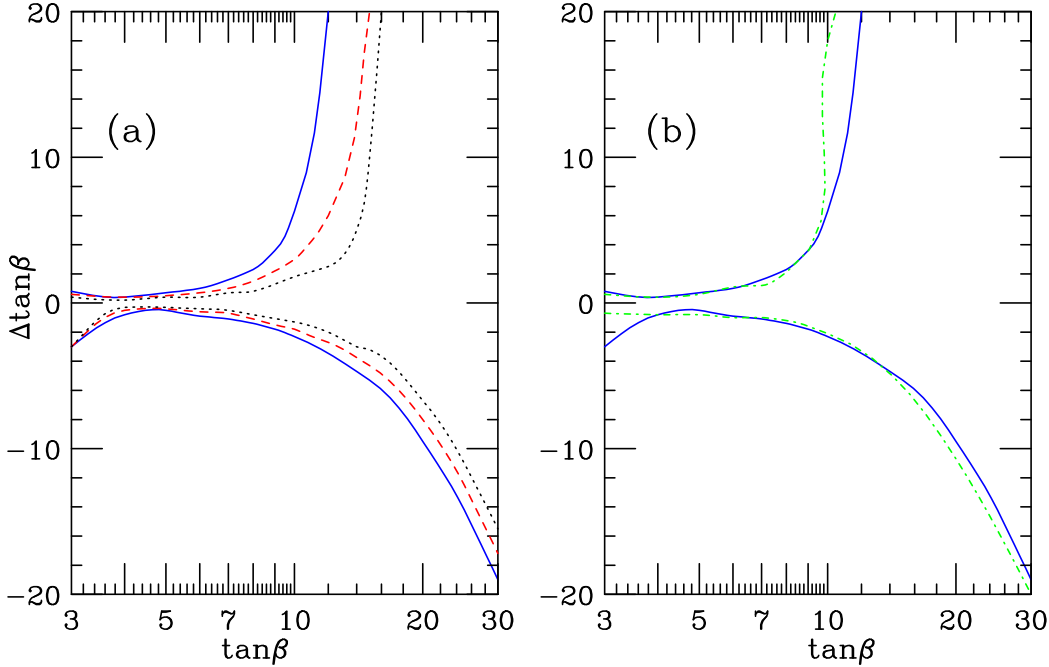,width=5.5in,angle=90}}
\caption{MSSM: Determination of $\tan\beta$ 
for $m_A^{}=200$ GeV at $\sqrt{s}$ = 500 GeV;
(a) $95\%$ C.L.~constraints on the $\tan\beta$ 
values for 100 fb$^{-1}$ (solid), 200 fb$^{-1}$ (dashes) 
and 500 fb$^{-1}$ (dotted), (b) Comparison of $95\%$ 
C.L.~constraints on $\tan\beta$ for 100 fb$^{-1}$ 
of our result (solid)
with that obtained from Ref.~\protect\cite{feng} (dot-dashed).
\label{fig:m200lim}}
\end{figure}

\begin{figure}[tbh]
\centerline{\psfig{file=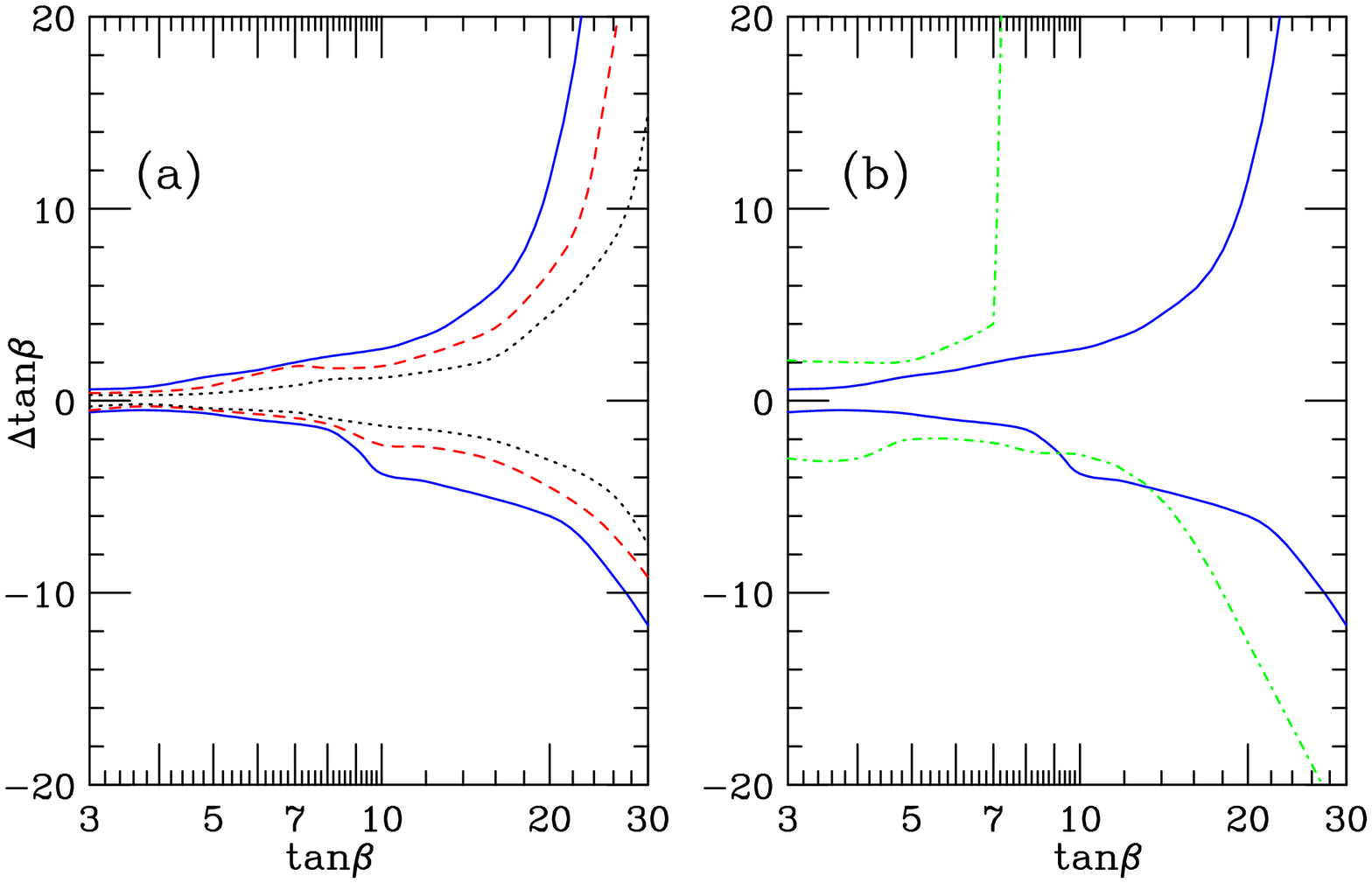,width=5.5in,angle=90}}
\caption{MSSM: Determination of $\tan\beta$ 
for $m_A^{}=400$ GeV at $\sqrt{s}$ = 1 TeV;
(a) $95\%$ C.L.~constraints on the $\tan\beta$ values
for 100 fb$^{-1}$ (solid), 200 fb$^{-1}$ (dashes) 
and 500 fb$^{-1}$ (dotted), (b) Comparison for $95\%$ 
C.L.~constraints on $\tan\beta$ for 100 fb$^{-1}$ 
of our result (solid)
with that obtained from Ref.~\protect\cite{feng} (dot-dashed).
\label{fig:m400lim}}
\end{figure}

We next consider the MSSM Scenario.
For the case with $m_A = 200$ GeV, the $A,H \to t\bar{t}$
decay channel is closed and we only make use of
the processes with $4b$ in the final state.  
With the lower Higgs masses, it is sufficient to
consider a linear collider with $\sqrt{s}$ = 500 GeV.  
The $95\%$ C.L.~constraints on the $\tan\beta$ determination
are show in Fig.~\ref{fig:m200lim}(a) for 100 fb$^{-1}$ 
(solid), 200 fb$^{-1}$ (dashes) and 500 fb$^{-1}$ (dotted). 
In Fig.~\ref{fig:m200lim}(b) we compare our result 
for 100 fb$^{-1}$ (solid) with that obtained by Feng and Moroi 
(dot-dashed) \cite{feng} and they are comparable.
For the case with $ m_A = 400$ GeV, the constraints on $\tan\beta$
values are shown in Fig.~\ref{fig:m400lim}, similar to
the previous figure. Since the $4t$ channel is available
in this case, the determination at low $\tan\beta$ is
significantly improved. For most values of $\tan\beta$,
in particular higher values,
we get more stringent constraints than the
results in \cite{feng}, indicating the
potential of better determination on $\tan\beta$ via
the neutral $H$ and $A$ channels under consideration.
We list our $\tan\beta$ constraints in Table~\ref{result} 
based on a 100 fb$^{-1}$ integrated luminosity
and compare with the values that we estimate from the results
by Feng and Moroi \cite{feng}, where a different statistical 
procedure of $\chi^2$ was adopted.
The results are largely comparable, but our constraints 
are somewhat tighter, especially for higher values of $\tan\beta$
as already seen in Fig.~\ref{fig:m400lim}(b).

\begin{table}[tbh]
\begin{center}
\begin{tabular}{c@{\hspace{1in}}c@{\hspace{1.5in}}c}
$\tan\beta$ & This analysis & Feng and Moroi\\
\hline
3 & $2.4< \tan\beta < 3.6$ & $\tan\beta < 5.2$\\
5 & $4.3 < \tan\beta < 6.3$ & $3.0 < \tan\beta < 6.0$\\
10 & $6.2 < \tan\beta < 12.7$ & $6.5 < \tan\beta$\\
20 & $14 < \tan\beta < 32$ & $7.5 < \tan\beta < 90$\\
30 & $18 < \tan\beta < 80$ & $8.0 < \tan\beta$\\
\end{tabular}
\vskip 0.2cm
\caption{Constraints on values of 
$\tan\beta$ by $95\%$ C.L.~statistical measurement
on the cross sections combining both $A$ and $H$ 
channels in the MSSM scenario; 
the results of Feng and Moroi shown here are estimated from 
the curves in Ref.~\protect\cite{feng} based on the
$t\bar b H^\pm$ process .
\label{result}}
\vskip -1cm
\end{center}
\end{table}

\section{Discussion and Conclusion}

For the mSUGRA scenario, $\tan\beta$ is essentially
the only variable after fixing the other soft
SUSY breaking parameters as in Eq.~(\ref{msugra}). 
The masses of the $H,A$ Higgs bosons decrease 
as $\tan\beta$ increases.  
Thus the corresponding Higgs branching fractions 
and the production cross sections
at a given energy increase with $\tan\beta$,
especially for large values.
This leads to possible accurate
determinations of $\tan\beta$ in mSUGRA (Fig.~\ref{fig:const}), 
for high values in particular.
In contrast, for the MSSM scenario the masses of $A$ and $H$ are 
independent of $\tan\beta$, and are kept fixed in the analyses.  
At large $\tan\beta$ values the decay branching fractions
and the production cross sections of $A$ or $H$ with 
$b\bar{b}$ reach a plateau in the MSSM. 
Consequently the determination of 
$\tan\beta$ in that range is less effective.  

There are other processes by which $\tan\beta$ 
may also be constrained:
(i) Chargino pair production in $e^+e^-$ collisions can provide
good measurements on $\tan\beta$ for low $\tan\beta$ 
values \cite{nlctanb,lctanb}; 
(ii) $\tilde{\tau}_L-\tilde{\tau}_R$ mixing
can be a sensitive probe of $\tan\beta$ \cite{Nojiri};
(iii) Gaugino production in $e\gamma$ collisions may provide information
on $\tan\beta$ \cite{han};
(iv) Kinematical distributions from the decay products of SUSY 
particles can be used to determine the $\tan\beta$ value
\cite{lhctanb};
(v) The magnetic dipole moment of the muon may be useful for 
$\tan\beta \agt 20$ if slepton masses $m_{\tilde{l}} \alt300$ 
GeV \cite{takeo}; 
(vi) The branching fractions of $H,A \to \tau\bar{\tau}$ may be 
useful to set a lower bound $\tan\beta \agt 10$ \cite{LHC,dieter}. 
The alternative methods in (i)-(iv) probe either $\sin\beta$
or $\cos\beta$; thus the sensitivity to $\tan\beta$ is degraded
at high values of $\tan\beta$. On the other hand, methods (v) and (vi)
are only effective for high values of $\tan\beta$.  In contrast, 
Higgs boson production processes under consideration and the
$t\bar b H^\pm$ process discussed in Ref.\cite{feng}
are direct probes 
of $\tan\beta$ with the complementary constraints from
$t\bar t$ and $b\bar b$ final states at low and high values of 
$\tan\beta$, respectively.

In summary, we studied heavy neutral Higgs boson production
in the minimal supersymmetric theories
at a linear collider with $\sqrt s = 0.5-1$ TeV with the 
expected integrated luminosities of $50-500\ \fbi$. 
The cross sections have a strong dependence on the fundamental
supersymmetry
parameter $\tan\beta$, and thus provide a good way to determine it.  
We considered the $4b$ and $4t$ final states which are sensitive
and complementary in determining $\tan\beta$. 
In the Supergravity scenario, the
sensitivity is particularly good for $\tan\beta \agt 10$
in comparison with other methods,
reaching a $15\%$ or better determination in a
$95\%$ C.L.~cross section measurement.
In the general MSSM scenario, the interplay
between the $4b$ and $4t$ channels results in a good
determination for $\tan\beta \alt 10$ (see Table~\ref{result}). 
For higher values
of $\tan\beta$ the sensitivity is weakened.
The accuracy of $\tan\beta$ determination is generally
sufficient to distinguish theories with a low value ($\sim 2$) from a
high value $(>30)$ and thus to provide information on
testing certain GUTs scenarios.

{\it Acknowledgments}:
We thank J. Feng and C. Kao for valuable discussions.  
This work was supported in part by
a DOE grant No. DE-FG02-95ER40896 and in part by 
the Wisconsin Alumni Research Foundation.

\end{document}